\documentclass{mn2e}
\newcommand{\kms}{\ifmmode {\rm km\ s}^{-1} \else km s$^{-1}$\ \fi}
\newcommand{\ergs}{\ifmmode {\rm erg\ s}^{-1} \else erg s$^{-1}$\ \fi}

\newcommand{\feii}{Fe {\sc ii}\ }
\newcommand{\mgii}{Mg {\sc ii}\ }
\newcommand{\civ}{C {\sc iv}\ }

\newcommand{\lb}{\ifmmode L_{\rm Bol} \else $L_{\rm Bol}$\ \fi}
\newcommand{\ledd}{\ifmmode L_{\rm Edd} \else $L_{\rm Edd}$\ \fi}
\newcommand{\lx}{\ifmmode L_{\rm 2-10keV} \else  $L_{\rm 2-10keV}$\ \fi}
\newcommand{\hb}{\ifmmode H\beta \else H$\beta$\ \fi}
\newcommand{\ha}{\ifmmode H\alpha \else H$\alpha$\ \fi}
\newcommand{\oiii}{[O {\sc iii}]\ }
\newcommand{\oii}{[O {\sc ii}]\ }
\newcommand{\nii}{[N {\sc ii}]\ }
\newcommand{\neiii}{[Ne {\sc iii}]\ }
\newcommand{\sii}{[S {\sc ii}]\ }
\newcommand{\mbh}{\ifmmode M_{\rm BH}  \else $M_{\rm BH}$\ \fi}
\newcommand{\lv}{\ifmmode \lambda L_{\lambda}(5100\AA) \else $\lambda L_{\lambda}(5100\AA)$\ \fi}
\newcommand{\msun}{M_{\odot}}
\newcommand{\mdot}{\ifmmode \dot{m} \else \dot{m} \fi }
\newcommand{\llog}{\ifmmode {\rm log} \else {\rm log} \fi }

\usepackage{graphicx}
\usepackage{color}
\usepackage{lscape}

\begin{document}
\title[The host of RE J1034+396]{The host galaxy of a narrow-line Seyfert 1 galaxy RE J1034+396 with X-ray quasi-periodic
oscillations}
\author[W. Bian \& K. Huang]
{Wei-Hao Bian\thanks{whbian@njnu.edu.cn} and Kai Huang \\
Department of Physics and Institute of Theoretical Physics, Nanjing
Normal University, Nanjing 210097, China\\} \maketitle

\begin{abstract}
Using simple stellar population synthesis, we model the bulge
stellar contribution in the optical spectrum of a narrow-line
Seyfert 1 galaxy RE J1034+396. We find that its bulge stellar
velocity dispersion is $67.7\pm 8$ \kms. The supermassive black hole
(SMBH) mass is about $(1-4)\times 10^6 \msun$ if it follows the
well-known $\mbh-\sigma_*$ relation found in quiescent galaxies. We
also derive the SMBH mass from the H$\beta$ second moment, which is
consistent with that from its bulge stellar velocity dispersion. The
SMBH mass of $(1-4)\times 10^6 \msun$ implies that the X-ray
quasi-periodic oscillation (QPO) of RE J1034+396 can be scaled to a
high-frequency QPO at 27-108 Hz found in Galactic black hole
binaries with a 10 $\msun$ black hole. With the mass distribution in
different age stellar populations, we find that the mean specific
star formation rate (SSFR) over past 0.1 Gyr is $0.0163\pm 0.0011$
$\rm Gyr^{-1}$, the stellar mass in the logarithm is $10.155\pm
0.06$ in units of solar mass, and the current star formation rate is
$0.23\pm 0.016~\msun ~\rm yr^{-1}$. RE J1034+396 does not follow the
relation between the Eddington ratio and the SSFR suggested by Chen
et al., although a larger scatter in their relation. We also suggest
that about 7.0\% of the total \ha luminosity and 50\% of the total
\oii luminosity come from the star formation process.
\end{abstract}

\begin{keywords}
galaxies: bulge --- galaxies: nuclei --- black hole physics
 --- galaxies: stellar content --- Galaxy:individual: RE J1034+396
\end{keywords}

\section{INTRODUCTION}
It is thought that active galactic nuclei (AGN) are scaled-up
versions of Galactic black hole binaries (BHBs; Gierlinski et al.
2008 and reference therein). The power spectrum of the X-ray
variability in BHBs shows the quasi-periodic oscillations of
0.01-450 Hz(QPOs; Rimillard \& McClintock 2006). It is believed that
we can also find the QPOs in AGN, on the similar behavior of
accretion flow around the black hole (BH). Comparing to a BH of 10
$\msun$ in BHBs, the typical high-frequency QPOs of 100 Hz would be
smaller by $10^6$ (i.e., $\sim 10^{-4}$ Hz) for a supermassive black
hole (\mbh; SMBH) of $10^7\msun$ if the QPO frequencies scale
inversely with the BH mass (Rimillard \& McClintock 2006).

With a long XMM-Newton observation (91 ks), Gierlinski et al. (2008)
detected a significant QPO signal ($\nu=2.7\times 10^{-4} \rm Hz$,
corresponding to a period of about 1 hour) for a nearby (z=0.043)
spiral active galaxy RE J1034+396 (J2000, RA=158.66082,
DEC=39.64119). It is optically classified as a narrow-line Seyfert 1
galaxy (NLS1), with similar small line width for the high-ionization
and low-ionization emission lines (Puchnarewicz et al. 1998), and a
very soft X-ray spectrum (Grupe et al. 2004; Casebeer et al. 2006).
NLS1s are thought to be a special subclass of AGN harboring
relatively small but growing SMBHs, compared with other broad-line
Seyfert 1 galaxies (BLS1s; e.g. Osterbrock \& Pogge 1985; Boller,
Brandt \& Fink 1996; Mathur 2000). By methods of the standard
accretion disk model fitting of spectral energy distribution (SED;
Puchnarewicz et al. 2001), the slim disk fitting of SED (Wang et
al., 1999; Wang \& Netzer 2003), the full width at half-maximum
(FWHM) of the H$\beta$ line, \oiii FWHM, and soft X-ray luminosity,
the SMBH mass in RE J1034+396 can not be determined very well, from
$6.3\times 10^5$ to $3.6\times 10^7\msun$ (Bian et al. 2004; Wang \&
Lu 2001). Its QPO type (high-frequency or low-frequency QPOs) can
not be uniquely identified (Gierlinski et al. 2008).

In the past two decades, there has been striking progress in finding
more reliable methods to calculate SMBHs masses in AGN through the
line width, $\Delta V$, of H$\beta$ (or H$\alpha$, \mgii, \civ) from
the broad line region (BLR) and the BLR size, $R_{\rm BLR}$ (e.g.,
Kaspi et al. 2000; McLure \& Dunlop 2004; Bian \& Zhao 2004;
Peterson et al. 2004; Greene \& Ho 2005b; Bian et al. 2008). There
are mainly two ways to parameterize the line widths of broad
emission lines, i.e., FWHM and the second moment ($\sigma_{line}$)
(e.g. Bian et al. 2008). The second moment of the broad components
of the H$\beta$ line provides a more precise measurement of the SMBH
mass because the H$\beta$ profile is non-Gaussian in NLS1s. It is
suggested that the mean value of the SMBH masses in NLS1s from the
H$\beta$ second moment is larger by about 0.50 dex than that from
FWHM (Bian et al. 2008). The well-known $\mbh - \sigma_{*}$ relation
of inactive galaxies (Tremaine et al. 2002) can also be used to
estimate the SMBH mass when the stellar velocity dispersion,
$\sigma_{*}$, is available (Kauffmann et al. 2003; Onken et al.
2004; Greene \& Ho 2006).

The X-ray QPOs frequency may depend on the mass and the spin of BH
(Rimillard \& McClintock 2006). The SMBH masses are important in the
study of QPOs in AGN. Here we use the the Sloan Digital Sky Survey
(SDSS; Abazajian et al. 2009) data of this object to determine its
SMBH mass by its $\sigma_*$ and H$\beta$ second moment instead of
H$\beta$ FWHM in Bian et al. (2004), as well as the star formation
history of its host bulge by the simple stellar population (SSP)
synthesis. All of the cosmological calculations in this paper assume
$H_{0}=70 \rm {~km ~s^ {-1}~Mpc^{-1}}$, $\Omega_{M}=0.3$, and
$\Omega_{\Lambda} = 0.7$.

\section{Data and Analysis}

\begin{center}
\begin{figure*}
\includegraphics[width=13cm]{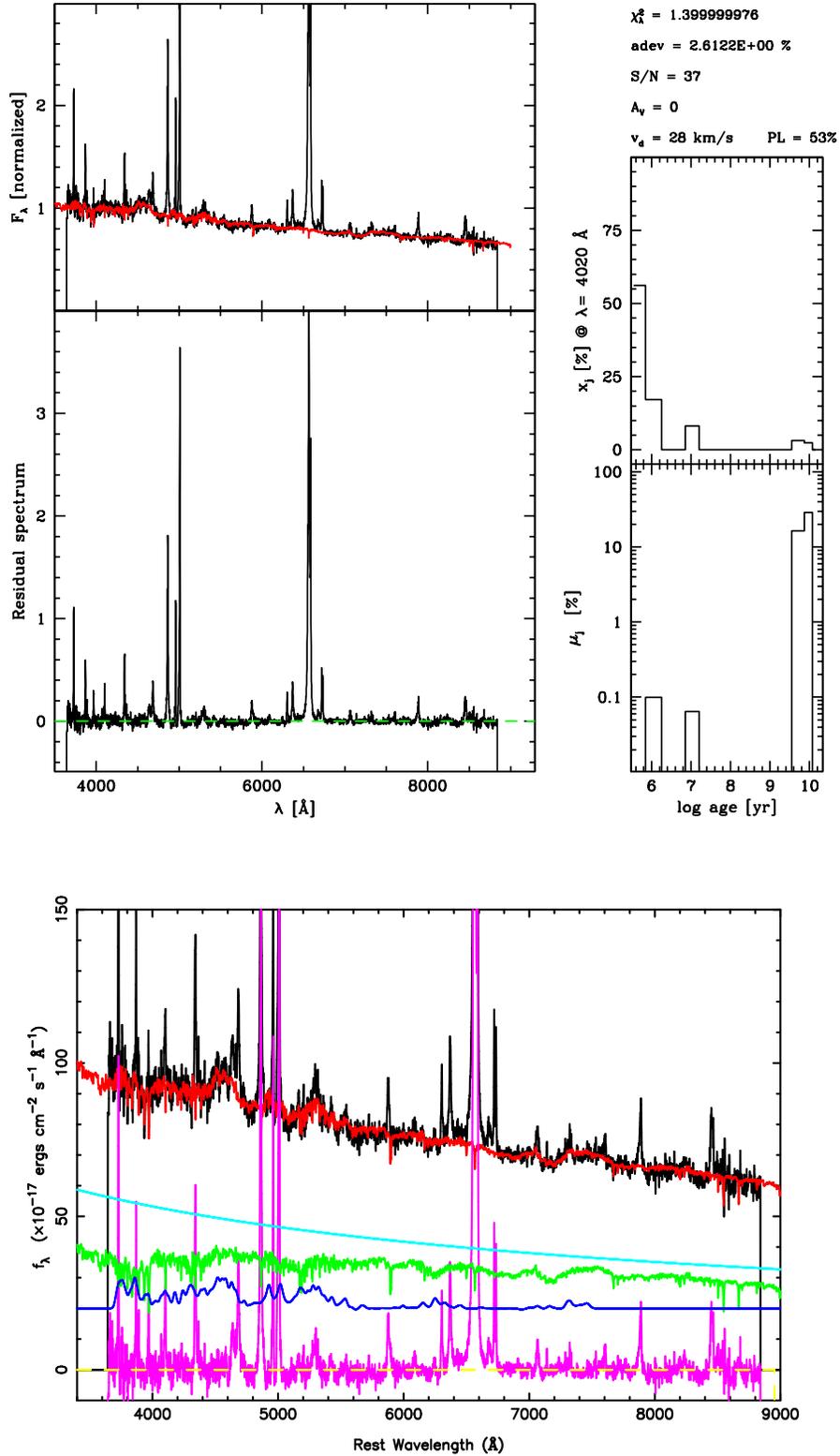}
\includegraphics[height=11cm,angle=-90]{f1b.eps}
\caption{Results of simple stellar population (SSP) synthesis for
the SDSS spectrum of RE J1034+396. Top panel: the left panels are
the observed and the synthesis spectra (black and red, respectively)
, and the residual spectrum. The panels on the right show the the
flux-fraction $x_j$ and mass-fraction $\mu_j$ versus the ages of the
simple stellar populations. Bottom panel: the black line is the
observed SDSS spectrum corrected for Galactic extinction in the rest
frame. The cyan line is the power-law continuum. The green line is
the host contribution. The blue line is the \feii spectrum. The
magenta line is the residue. The red line is the composition of the
host contribution, \feii lines and the power-law continuum.}
\end{figure*}
\end{center}

The SDSS used a dedicated 2.5-m wide-field telescope at Apache Point
Observatory near Sacramento Peak in Southern New Mexico to conduct
an imaging and spectroscopic survey for about 1/4 sky area. The SDSS
data have been made public in a series of yearly data releases and
the present is the Seventh Data Release (DR7), including data up to
the end of SDSS-II in July, 2008.


The optical spectrum of RE J1034+396 is downloaded from SDSS DR7.
Its spectrum covers the wavelength range of 3800-9200 \AA\ with a
spectral resolution of $1800 < R < 2100$. At its redshift of 0.043,
the projected fiber aperture diameter of 3" is about 2.4 kpc,
containing most light from its bulge. With nuclei spectrum
superimposed on it, the stellar absorption features in its SDSS
spectrum provide us the possibility to investigate the property of
its host bulge, including the bulge stellar velocity dispersion, the
stellar mass, and the star formation history. We did not apply
aperture correction to the stellar velocity dispersion because this
effect can be omitted for $z<0.3$ (Bernardi et al. 2003).

We outline our steps to do the SDSS spectral analysis. (1) We use
SSP synthesis (STARLIGHT; Cid Fernandes et al. 2005) to model the
stellar contribution in the Galactic extinction-corrected spectrum
in the rest frame (Cid Fernandes et al. 2005; Bian et al. 2006,
2007, 2008). The Galactic extinction law of Cardelli, Clayton \&
Mathis (1989) with $R_{V} = 3.1$ is adopted, which is also used for
the host extinction with V-band extinction (from 0 mag to 5.0 mag).
We use 45 default templates in Cid Fernandes et al. (2005), which
are calculated from the model of Bruzual \& Charlot (2003). The
linear combination of 45 templates is used to represent the host
bulge spectrum. These 45 templates comprise 15 ages, $t=$ 0.001,
0.00316, 0.00501, 0.01, 0.02512, 0.04, 0.10152, 0.28612, 0.64054,
0.90479, 1.434, 2.5, 5, 11 and 13 Gyr, and three metallicities, $Z=$
0.2, 1 and 2.5 $Z_\odot$ (Cid Fernandes et al. 2005). At the same
time as the SSP fit, we add a power-law component in the code to
represent the AGN continuum emission, and an optical \feii template
from the prototype NLS1 I ZW 1 (Boroson \& Green 1992) to model the
\feii emission. We exclude the AGN mission lines, such as H Balmer
lines, \oii$\lambda$3727, \neiii$\lambda$3869, \oiii$\lambda
\lambda$4959, 5007, \nii$\lambda \lambda$6548, 6583, \sii$\lambda
\lambda$6717, 6731.

The synthetic spectrum is built using the following equation,
\begin{equation}
M_\lambda = M_{\lambda_0}
   \left[
   \sum_{j=1}^{N_*} x_{\rm j} b_{\rm j,\lambda} r_\lambda
   \otimes G(v_0,v_d)+ x_{\rm fe}\otimes G(v_{\rm fe},\sigma_{\rm fe})
   \right]
\end{equation}
where $b_{\rm j,\lambda}$ is the $j^{\rm th}$ template normalized at
$\lambda_0=4020$ \AA, $x_j$ is the flux fraction at 4020 \AA,
$M_{\lambda_0}$ is the synthetic flux 4020 \AA, $r_{\lambda}\equiv
10^{-0.4(A_{\lambda}-A_{\lambda 0})}$ is the reddening term by
V-band extinction $A_V$, and $G(v_0,v_d)$ is the line-of-sight
stellar velocity distribution, modeled as a Gaussian centered at
velocity $v_0$ and broadened by the velocity dispersion $v_d$. Due
to different velocity dispersion in the stellar lines and \feii
lines, we use another line-of-sight \feii velocity distribution
$G(v_{\rm fe}, \sigma_ {\rm fe})$ for the \feii emission. $x_{\rm
fe}$ is the \feii flux-fraction at 4020\AA. The line-of-sight \feii
velocity distribution $G(v_{\rm fe},\sigma_{\rm fe})$ is also
modeled as a Gaussian centered at velocity $v_{\rm fe}$ and
broadened by the velocity dispersion $\sigma_{\rm fe}$. We first fit
the \feii lines and the continuum in the fitting windows. And we
find that $\sigma_{\rm fe}=411\pm 46 \kms$. $\sigma_{\rm fe}$ is
fixed by 411 km/s in the fitting of SSP and \feii. When we change
$\sigma_{\rm fe}$, the SSP and \feii fitting results do not change,
considering the errors. We also exclude the \feii emissions and do
the SSP fit, considering the errors, the SSP results do not change.
The best fit is reached by minimizing reduced $\chi^2$,
\begin{equation}
\chi^2(x,M_{\lambda_0},A_{\rm V},v_0,v_d) =
   \sum_{\lambda=1}^{N_\lambda}
   \left[
   \left(O_\lambda - M_\lambda \right) w_\lambda
   \right]^2/N
\end{equation}
where the weighted spectrum $w_\lambda$ is defined as the noise
associated with each spectral bin as reported by the SDSS pipeline
output, N is the total unmasked pixels. $\chi^2$ is calculated by
the difference between observed spectrum and the model spectrum in
the fitting of SSP and \feii. For RE J1034+396, the best fit of SSP
and \feii gives $\chi^2=1.4$ (Fig 1). We find that $A_{V}=0$, and
the host extinction can be neglected (Fig 1). Through above spectral
synthesis, we can obtain some parameters, such as bulge velocity
dispersion $v_d$, flux-fraction $x_j$, mass-fraction $\mu_j$,
stellar mass $M_*$. The mass-flux ratio can be found in STARLIGHT
manual on the web site http://www.starlight.ufsc.br/. The results
are shown in Fig. 1 and Table 1.

With the simulation, it is suggested that the uncertainty of SSP
results can be given by the effective starlight signal-to-noise
(S/N) at 4020 \AA\ (Cid Fernandes et al. 2005; Bian et al. 2007).
The S/N at 4020 \AA\ is 37 and the starlight fraction at 4020 \AA\
is about 43\% (Table 1). Therefore the effective starlight S/N at
4020 \AA\ is about 16, corresponding to an uncertainty of 8 \kms for
the velocity dispersion, 7\% for the mass-fraction, 8\% for the
flux-fraction, 0.06 dex for stellar mass $log M_*$ (Cid Fernandes et
al. 2005; Bian et al. 2007). We adopt 10\% as the uncertainty of
host bulge flux fraction.

(2) Considering the broad wing in the H$\beta$ line profile, two
broad components are used to model broad H$\beta$ profile from BLRs.
The H$\beta$ line from the narrow line regions (NLRs) has the same
profile as the \oiii line from NLRs. Two components are used to
model the asymmetric \oiii line profile, and two components are used
to model H$\beta$ line profile from NLRs (Bian et al. 2008).
Therefore, total four Gaussians are used to model the H$\beta$ line
profile, and two sets of two Gaussians are used to model the
\oiii$\lambda \lambda 4959, 5007$ lines. As the same as the H$\beta$
profile, four Gaussians are used to model the H$\alpha$ profile (two
broad components from BLRs and two narrow components from NLRs), and
two sets of two Gaussian are used to model the \nii $\lambda \lambda
6548, 6583$ lines. We take the same line width for each
corresponding component of \oiii $\lambda \lambda 4959, 5007$ and
H$\beta$ from NLRs, fix the flux ratio of \oiii  $\lambda 4959$ to
\oiii $\lambda 5007$ to be 1:3, and set the wavelength separation to
the laboratory value. We take the same line width for each
corresponding component of \nii $\lambda \lambda 6548, 6583$ and
H$\alpha$ from BLRs, fix the flux ratio of \nii $\lambda 6548$ to
\nii $\lambda 6583$ to be 1:3, and set the wavelength separation to
the laboratory value. For RE J1034+396, the best fits of the \hb,
\ha lines give reduced $\chi^2=2.2, \chi^2=2.14$, respectively. The
results are shown in Fig. 2 and Table 2. The third component of the
H$\beta$ line from NLRs is weaker, the results do not change if we
remove this weaker component.

\begin{center}
\begin{figure}
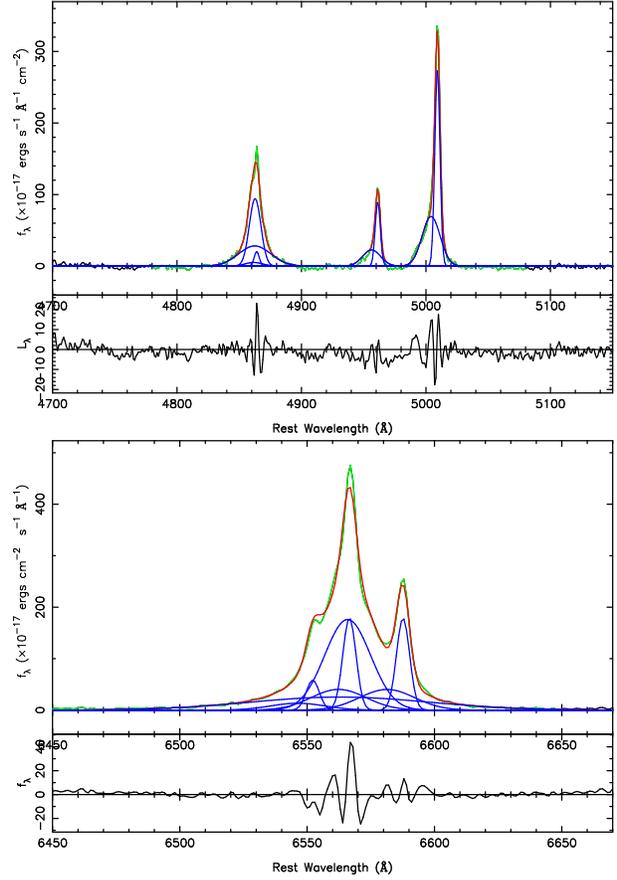

\includegraphics[height=8cm,angle=-90]{f2a.eps}
\includegraphics[height=8cm,angle=-90]{f2b.eps}
\caption{An example of the line fit for RE J1034+396. Top panel: the
line fit for \hb and \oiii line. The black line is the original
spectrum after Galactic-extinction correction, the starlight
subtraction, the power-law continuum and the \feii subtraction in
the rest frame. The residual is shown at the bottom. The multiple
Gaussian components are in blue and the sum of them is in red. The
fitting window is in green. Bottom panel: the line fit for \ha and
\nii line. The line colors are the same to that in top panel.}
\end{figure}
\end{center}

\section{Result and Discussion}

\begin{table}
\caption{The host bulge properties of RE J1034+396}
\begin{tabular}{llllllllll}
\hline \hline
$\sigma_*(\kms)$  &  $log M_*(\msun)$                       & $SSFR(\rm Gyr^{-1})$             & $f_{\rm host}$  \\
(1) & (2)& (3)& (4) \\
\hline
$67.7\pm 8$   &  $10.155\pm 0.06$ & $0.0163\pm 0.0011$ & $43\pm 4$\% \\
\hline
\end{tabular}

Note. Col(1): the host bulge stellar velocity dispersion; Col(2):
the stellar mass; Col(3): the host bulge specific star formation
rate; Col(4): the host bulge flux fraction at 4020 $\AA$ with 10\%
uncertainty.
\end{table}

\begin{table}
\caption{The emission lines properties of RE J1034+396}
\begin{tabular}{cllccc}
\hline \hline

    & FWHM$^a$  & flux$^b$ \\
\hline
\hb & $1690\pm 296$ &  $1.1\pm 0.84$ \\
    & $617 \pm 155$  &  $0.88\pm 0.73$ \\
    & $1022\pm 4592^{n}$ &  $0.10\pm 0.63$ \\
    & $292 \pm 114^{n}$  &  $0.12\pm 0.13$ \\
\oiii & $1023\pm18^{n}$ &  $1.26\pm 0.03$ \\
      & $292\pm 5^{n}$  &  $1.43\pm 0.03$ \\
\ha  & $969\pm 552 $ & $3.98\pm 3.15$ \\
     & $3804\pm 1100$ & $2.25 \pm 0.39$ \\
     & $272 \pm 25^{n}$  &  $1.16\pm0.25$ \\
     & $1162 \pm 275^{n}$  &  $1.09\pm2.75$ \\
\nii $\lambda 6583$ & $278 \pm 80^{n}$ & $1.16\pm 0.08$ \\
                    & $1162 \pm 287^{n}$ & $1.1\pm 2.81$ \\
\feii & $967\pm 108 $ & $1.76\pm 0.2^c$ \\

\hline

\end{tabular}

Note. $^a$: FWHM of a Gaussian profile in units of \kms. $^b$: in
units of $10^{-14}~\rm erg~s^{-1}~ cm^{-2}$. $^c$: \feii flux
between 4434 \AA\ and 4684 \AA. $^n$: the components from NLRs. The
second weaker broad component of H$\alpha$ is probably due to the
contribution from the continuum subtraction.

\end{table}

\subsection{Mass}

Firstly, we derive the mass from the bulge stellar velocity
dispersion, $\sigma_*$. By SSP method, the measured bulge stellar
velocity dispersion $v_d$ is $28\pm 8$ \kms. Considering the
resolutions of the SDSS spectra and the template spectra, we have to
do these corrections by the following formula {\footnote{
http://www.starlight.ufsc.br/}},
\begin{equation}
\sigma_*=\sqrt{v_d^2+\sigma_{\rm temp}^2-\sigma_{\rm inst}^2}
\end{equation}
where the SDSS spectral resolution $\sigma_{\rm inst}$ is adopted as
60 \kms (Greene \& Ho 2005a), the template spectral resolution
$\sigma_{\rm temp}$ is adopted as 86 \kms (Cid Fernandes et al.
2005). Assuming that the errors on $\sigma_{\rm inst}$ and
$\sigma_{\rm temp}$ are negligible, the error on $\sigma_*$ will be
the same as that on $v_d$.  $v_d=28 \pm 8 \kms$ leads to
$\sigma_*=67.7\pm 8\kms$. Using the $\mbh - \sigma_{*}$ relation for
quiescent galaxies, $\mbh (\sigma_*) = 10^{8.13}[\sigma_{*}/(200 \
\rm km s^{-1})]^{4.02} ~~\msun $ (Tremaine et al. 2002), we derive
the SMBH mass is $(1.7^{+1.0}_{-0.7}) \times 10^{6}~\msun$ for
$\sigma_*=67.7\pm 8\kms$. The uncertainty of 8 \kms for $\sigma_*$
would lead to an error of 0.1 dex for \llog \mbh. The total
uncertainty is about 0.32 dex considering the error of 0.3 dex from
the $\mbh - \sigma_{*}$ relation (Tremaine et al. 2002). Therefore,
the SMBH mass of RE J1034+396 derived from $\sigma_*$ is about
$(1-4)\times 10^6 \msun$.

Secondly, we derive the mass from the broad \hb second moment
$\sigma_{\rm H\beta}$ and the empirical $R_{\rm BLR}-\lv$ relation
(Kaspi et al. 2000, 2005; Bentz et al. 2006; Bian et al. 2008). With
the power-law component, we can derive the nuclear flux at 5100 \AA\
is $45\times 10^{-17}\rm ~erg~s^{-1}~ cm^{-2}$, and \lv is
$5.6\times 10^{42} ~\ergs$. The first moment of the line profile
$P(\lambda)$ is
\begin{equation}
\lambda_{0}  = \frac{\int \lambda P(\lambda) d\lambda}{\int
P(\lambda) d\lambda},
\end{equation}
The second moment of the line profile $P(\lambda)$ is
\begin{equation}
\sigma^2_{line}=\frac{\int \lambda^2 P(\lambda) d\lambda}{\int
P(\lambda) d\lambda}-\lambda_{0}^2.
\end{equation}
For a Gaussian line profile, the ratio of FWHM to the second
momentum is $\rm FWHM/\sigma_{\rm line}=\sqrt{8ln2} \approx 2.35$;
while for a Lorentzian profile, $\sigma_{\rm line} \rightarrow
\infty$. For RE J1034+396, the H$\beta$ second momentum $\sigma_{\rm
H\beta}$ is calculated from the reconstructed \hb profile of two
broad H$\beta$ components from BLRs in step 2 of spectral analysis,
and $\sigma_{\rm H\beta}$ is $577\pm144$ \kms. From the
reconstructed H$\beta$ profile from BLRs, FWHM is $802\pm 200 \kms$.
The SMBH mass can be calculated from the H$\beta$ second moment by
the following equation(Bian et al. 2008),
\begin{eqnarray}
M_{\rm BH} = f\frac{R_{\rm BLR} \Delta V^2}{G}~~~~~~~~~~~~~~~~~~~~~~~~~~~~~~~~~~~~~~~~~~~~\nonumber  \\
=f \times 7.629\times
10^6[\frac{\lv}{10^{44}\ergs}]^{0.518}[\frac{\sigma_{\rm \hb}}{1000
\kms}]^2 \msun
\end{eqnarray}
where $R_{\rm BLR}=39.08\times [\lv/(10^{44}\ergs)]^{0.518}$
light-days (Bentz et al. 2006), and $f=3.85$ (Collin et al. 2006).
With $\sigma_{\rm \hb}=577\pm 144 \kms$ and $f=3.85$, the estimated
SMBH mass is $(0.7-3.4)\times 10^6 \msun$. The uncertainty of the
mass calculation from the \hb line is mainly from the systematic
uncertainties, up to about 0.5 dex, which is due to the unknown
kinematics and geometry in BLRs (e.g. Krolik 2001; Peterson et al.
2004). Grupe et al. (2004) gave \lv as $1.5\times 10^{43} ~\ergs$.
Considering the half light contribution from the host bulge, the
nuclear \lv is consistent with ours. They also gave the \hb FWHM as
$700\pm 110$ \kms. When using this FWHM value to calculate the mass,
the mass would be smaller by 0.5 dex than that from the $\sigma_{\rm
\hb}$ (Bian et al. 2004, 2008). The new mass from the H$\beta$
second moment is consistent with that from the $\mbh - \sigma_{*}$
relation. Hereafter, we adopt $(1-4)\times 10^6 \msun$ as its SMBH
mass. This SMBH mass value of RE J1034+396 is consistent with that
derived from the accretion disk fitting of its SED (Puchnarewicz et
al., 2001; Wang \& Netzer 2003).

It is suggested that the gaseous kinematics of NLRs be primarily
governed by the bulge gravitational potential. Using the linewidth
of \oiii or \nii to trace the $\sigma_*$, we find that, from narrow
\oiii component, $\sigma_{\rm [\rm OIII]}=124$ \kms (see Table 2).
It would lead to a mass of $2\times 10^7 \msun$, which is larger
than mass from $\sigma_*$ and $\sigma_{\rm \hb}$ by an order of
magnitude. For RE J1034+396, $\sigma_{\rm [\rm O III]}$ cannot be
used to trace $\sigma_*$.

\subsection{High-frequency QPOs and BH spin}
QPOs can be classified as low-frequency QPOs (roughly 0.1-30 Hz) and
high-frequency QPOs (roughly 40-450 Hz, Rimillard \& McClintock
2006). Assuming QPO frequency ($\nu_{\rm BHB}$) scales inversely
with the BH mass ($M_{\rm BHB}$) in BHBs, $\mbh/M_{\rm BHB}=\nu_{\rm
BHB}/\nu_{\rm BH}$, where $\nu_{\rm BH}$ is the QPO frequency for
SMBH in AGN. By $\nu_{\rm BHB}=\nu_{\rm BH} \times (\mbh/M_{\rm
BHB})$, the QPO frequency of $2.7\times 10^{-4}$ Hz in RE J1034+396
(its SMBH mass of $2\times 10^6 \msun$ with an uncertain of a factor
2) corresponds a frequency of 27-108 Hz in Galactic BHBs with a 10
$\msun$ BH. Our result of 27-108 Hz suggests that the QPO found in
RE J1034+396 belongs to a high-frequency QPO. Larger estimated SMBH
mass, smaller adopted BHBs mass would make this frequency value
larger.

It is believed that the high-frequency QPOs may depend only on the
mass and the spin of the BH. This dependence can be expected for
coordinate frequencies (see Merloni et al. 1999) or for disk
oscillation modes in the inner accretion disk (see Kato 2001). For a
Schwarzchild BH, the innermost stable circular orbit (ISCO)
corresponds a maximum orbit frequency as $\nu_{\rm ISCO}=2200
(\mbh/\msun)^{-1} \rm Hz$. For an extreme Kerr BH, $\nu_{\rm
ISCO}=16150 (\mbh/\msun)^{-1} \rm Hz$ (Rimillard \& McClintock
2006). For three BHBs with high-frequency QPOs and well-constrained
BH masses, Rimillard \& McClintock (2006) suggested an empirical
frequency - mass relation for the high-frequency QPOs,
$\nu_0=931(\mbh/\msun)^{-1}$ Hz. With the SMBH mass of $2\times 10^6
\msun$ in RE J1034+396, the corresponding frequencies derived from
the ISCO of a Schwarzchild BH, the ISCO of an extreme Kerr BH, and
the empirical frequency - mass relation are $11.0\times 10^{-4}$,
$8.0\times 10^{-4}$, $4.7\times 10^{-4}$ Hz, respectively.
Considering an uncertainty of a factor of 2 in the SMBH mass and
other uncertainty in empirical frequency - mass relation for the
high-frequency QPOs (e.g., the BH spin), the double or triple value
of the fundamental frequency $4.7\times 10^{-4}$ Hz dose not deviate
much from the observed frequency of $2.7\times 10^{-4}$ Hz in RE
J1034+396. It implies that the QPOs phenomenon has the same origin
in BHBs and AGN (e.g., Gierlinski 2008).

\subsection{Star formation}
The widely employed method to measure the current star formation
rates (SFR) is by means of the ultraviolet continuum, the emission
lines of \ha, \oii, et al. (Kennicutt 1998). For a sample of 82302
star formation galaxies from SDSS, Asari et al. (2007) investigated
the current SFR derived from the SSPs synthesis and found that it is
consistent with that from the H$\alpha$ SFR indicator (see their
Fig. 6). The current specific SFR (SSFR) is defined by the mean SSFR
over the past 0.1Gyr, i.e., first 21 of the 45 templates described
in section 2 with age less then 0.1 Gyr (Asari et al. 2007; Chen et
al. 2009),
\begin{equation}
{\rm SSFR(t<0.1Gyr)}=\frac{\rm
SFR}{M_*}=\frac{\sum^{21}_{j=1}\mu_j}{\rm 0.1Gyr}
\end{equation}
For RE J1034+396, current $\rm SSFR(t<0.1Gyr)$ is $0.0163\pm 0.0011$
$\rm Gyr^{-1}$, $\llog M_*$ is $10.155\pm 0.06 \msun$, and current
SFR is $0.23\pm 0.016~\msun yr^{-1}$. Using the SFR formula from
Kennicutt et al. (1998), this current SFR of $0.23\pm 0.016~\msun
yr^{-1}$ corresponds to the \ha luminosity as $(2.9\pm 0.2) \times
10^{40}$ \ergs or the \oii luminosity as $(1.6\pm 0.1) \times
10^{40}$ \ergs coming from the starburst process. From the SDSS
spectrum, we find that total \ha luminosity and total \oii
luminosity is $4.4 \times 10^{41}$ \ergs, $3.4 \times 10^{40}$
\ergs. Therefore, about 6.1\%-7.0\% of the total \ha luminosity and
44\%-50\% of the total \oii luminosity come from the star formation
(Ho 2005).

We also use 150 templates instead of 45 templates in SSP synthesis
to model the stelar absorption contribution, like in Asari et al.
(2007). We found that $\sigma_* = 72.4 \kms$, $\llog M_* = 10.244$,
$\rm SSFR(t<0.1Gyr) = 0.02786 Gyr^{-1}$. Using 150 templates, the
$\sigma_*$ and $\llog M_*$ are consistent with that using 45
templates, the stellar flux fraction is 47\%, and the SSFR would be
larger by 70\%.

Using type-II AGN sample from MPA/JHU catalog (Kauffmann et al.
2003), Chen et al. (2009) found a correlation between the Eddington
ratio $\lambda$ (the ratio of the bolometric luminosity, $L_{\rm
bol}$, to the Eddington luminosity, $L_{\rm Edd}$) and the mean
SSFR, $\llog \lambda=(-0.73\pm 0.01)+(1.5\pm 0.01)\llog(\rm
SSFR/Gyr^{-1})$, suggesting that supernova explosions play a role in
the transportation of gas to galactic centers. Considering that the
difference between type II AGN and type I AGN is due to the
orientation of the line of sight, type I AGN would also follow this
$\llog \lambda - \llog \rm SSFR$ relation. Here we calculate the
Eddington ratio for RE J1034+396. Using $L_{\rm bol}=9\times \lv$
(Kaspi et al. 2000), $L_{\rm Edd}=1.26 \times 10^{38} (M_{\rm
bh}/\msun) \ \ergs$, and $\mbh=2\times 10^6 \msun$, we find that the
Eddington ratio is 0.2. Considering the uncertainty of a factor of 2
in \mbh, the uncertainty of the Eddington ratio is a factor of 2,
i.e., $\llog \lambda=-1\sim -0.4$. Using the formula of (3) in Chen
et al. (2009), the SSFR of $0.0163\pm 0.001$ $\rm Gyr^{-1}$ leads to
$\llog \lambda$ as $-3.45 \sim -3.37$, which deviates much from the
value of $-1\sim -0.4$. The deviation of $\llog \lambda$ is $2.37
\sim 3.05$ in the $\llog \lambda - \llog \rm SSFR$ diagram of Chen
et al. (2009). Considering most of the bolometric luminosity is
emitted in the UV and soft X-ray band, Grupe et al. (2004) estimated
the bolometric luminosity from a combined power-law model fit with
exponential cutoff to the optical-UV data and a power law to the
soft X-ray data. They found $L_{\rm bol}\approx 22 \times \lv$,
which is consistent with its hot big blue bump. If that is the case,
$\llog \lambda$ would be larger by 0.39. Therefore, RE J103 4+396
does not follow the relation between the Eddington ratio and SSFR
found by Chen et al. (2009), although a large scatter in their
relation. The reason for the deviation would be explored in other
place.

\section{Conclusions}
The host bulge of a NLS1 RE J1034+396 with X-ray quasi-periodic
oscillations is investigated through its optical spectrum from SDSS
DR7. The main conclusions can be summarized as follows: (1) The host
bulge flux contribution at 4020 \AA\ is about $(43\pm 4)\%$ in the
SDSS spectrum and the bulge stellar velocity dispersion $\sigma_*$
is $67.7\pm 8$ \kms. Considering the scatter in $\mbh - \sigma_{*}$
relation, the SMBH mass from $\mbh - \sigma_{*}$ relation is
$(1-4)\times 10^6 \msun$. (2) Using multi-Gaussians to model the
\hb, \oiii, \ha, and \nii emission lines, we find that the mass from
$\sigma_{\rm \hb}$ and \lv is $(0.7\pm 3.4)\times 10^6 ~\msun$,
consistent with that from $\sigma_*$. However, the \oiii velocity
dispersion is $124\pm 2$ \kms, about a double of $\sigma_*$. (3) The
SMBH mass of about $(1-4)\times 10^6 \msun$ implies that the QPO of
$2.7\times 10^{-4}Hz$ found in RE J1034+396 can be scaled to a
high-frequency QPO at about 27-108 Hz found in Galactic BHBs with a
10 $\msun$ BH. (4) From the SSPs synthesis, we find that current
$\rm SSFR(t<0.1Gyr)$ is $0.0163\pm 0.0011 \rm Gyr^{-1}$, $\llog
(M_*/\msun)$ is $10.155\pm 0.06$, and current SFR is $0.23\pm
0.016~\msun \rm yr^{-1}$. RE J1034+396 does not follow the relation
between the Eddington ratio and SSFR. About 6.1\%-7.0\% of the total
\ha luminosity and 44\%-50\% of the total \oii luminosity come from
the star formation process.

\section*{ACKNOWLEDGMENTS}
This work has been supported by the NSFC (Nos. 10873010 and 1073301)
and National Basic Research Program of China - the 973 Program
(Grant No. 2009CB824800). We are very grateful to the anonymous
referee for her/his instructive comments which significantly
improved the content of the paper. We thank J. M. wang, S.N. Zhang,
C. Hu, and Q. S., Gu for their useful discussions.


\end{document}